\documentclass{article}
\usepackage{spconf,amsmath,graphicx}

\title{ANALYSIS OF TWITTER TRAFFIC BASED ON RENEWAL DENSITIES}
\name{Javier Esteban, Antonio Ortega, Sean McPherson and Maheswaran Sathiamoorthy}
\address{Ming Hsieh Department of Electrical Engineering\\
University of Southern California, Los Angeles, CA, 90089-2564}
\begin{document}
\ninept
\maketitle
\begin{abstract}
In this paper we propose a novel approach for Twitter traffic analysis based on renewal theory. Even though twitter datasets are of increasing interest to researchers, extracting information from message timing remains somewhat unexplored. Our approach, extending our prior work on anomaly detection, makes it possible to characterize levels of correlation within a message stream, thus assessing how much interaction there is between those posting messages. Moreover, our method enables us to detect the presence of periodic traffic, which is useful to determine whether there is spam in the message stream. Because our proposed techniques only make use of timing information and are amenable to downsampling, they can be used as low complexity tools for data analysis. 
\end{abstract}
\begin{keywords}
Twitter, Discrete event systems, renewal density, spam 
\end{keywords}
\section{Introduction}
\label{sec:intro}

Twitter is a micro-blogging service that allows users to post messages of up to 140 characters (\emph{tweets}), which can contain text and URLs. It is also possible to resend a tweet posted by another user (\emph{retweet}), mention a user in the tweet (\emph{@username}) or tag a message (\emph{hashtags} with a tag symbol '\#'). This service has attracted millions of users in a short period of time. Its lightweight nature and its ability to reach many users very rapidly has helped it to have a significant role in some recent social events. Because users interact with each other on a wide variety of topics, it is also seen as a potentially important source of information, e.g., to spot emerging trends. This has led to increased interest in mining twitter streams for information. At the same time, the sheer number of messages transmitted makes this type of data mining very challenging. 

One of the main research areas related to Twitter is spam detection, where two main kinds of approaches have been proposed. Graph based methods use users' connections to detect spamming accounts. Examples of those methods are: PageRank, Hits, NodeRanking, TunkRank and Twitter Rank \cite{Brenes}. Alternatively, content based methods focus on the tweets' content and use classification techniques such as tree based methods \cite{Chandramouli} and support vector machines (SVM) \cite{Wang}\cite{Benevenuto}. Spammers may use different strategies in order to avoid Twitter's spam control, such as duplicating tweets with only a few changes (e.g., a mention, hashtag or URL \cite{Wang}). Spammers also use trend topics to reach the largest number of users, even though their message is not related with the trend topic. Thus, when there is no relation between the trend topic and the message, it may be classified as spam. In \cite{Benevenuto} the authors focus on spammers that use unrelated URLs within the text and retweet legitimate tweets changing them into illegitimate tweets obfuscated under an URL shortener. The age of the account and the posting rate during the beginning period of the account are also taken into consideration.
In addition to spam detection, researchers are also analyzing twitter streams in order to extract different types of information about messages or users. Examples include user classification, with an aim to classify user types and intentions, as well as geographical usage distribution \cite{Horn}. Recommendation systems such as \emph{Buzzer} or \emph{zerozero88}  and knowledge acquisition targets, such as  \emph{homophily} and \emph{Tweetonomies} have also been proposed \cite{Horn}. A measure of a user influence is given in \cite{Cha} and \cite{Lee}, while \cite{Sharifi} proposes an automatic summarization into topics. Finally, text classification labels natural language into a thematic categories. There are many systems, being \emph{Term-Document Matrix}, \emph{Vector space model} or \emph{n-grams} some of them \cite{Horn}.

We note that most of the techniques considered to date do not make use of tweet timing information, or more specifically, they do not study how the relative timing of tweets can provide useful information about the corresponding message streams. The key novelty of our work is to use timing information to detect the presence of spam (since spam messages are more likely to be posted in regular intervals) and to infer the level of interaction between users. To the best of our knowledge, we are the first to consider twitter message timing information for these purposes. In addition to uncovering new types of information from twitter data, timing based techniques have the advantage of being relatively low cost and of being amenable to efficient data downsampling \cite{McPherson}. 

We extend the work in \cite{McPherson}, where renewal theory was used to detect low-rate periodic events in Internet traffic. The renewal density (rd) is estimated from the empirical data using two different approaches: i) by combining histograms estimating pdfs of different orders \cite{Markovich} and 2) by using a convolution to compute the higher order inter-arrival pdfs taking the first order pdf as a starting point. As in \cite{McPherson}, we identify periodic events in an inter-arrival sequence, with the goal of detecting the presence of spam. Moreover, we propose using the renewal theory framework for a novel target, namely, characterizing the amount of memory in the sequence of inter-arrival times, with the goal of determining how much interaction there is between users posting about a certain topic. The main intuition here is that interaction between users (messages sent in response to other messages) will lead to message inter-arrival times that are more correlated.  Our initial results are promising.  After analyzing three different classes of keywords, we show that different levels of correlation can be found in the traffic, and that these correspond to the level of interaction that can be expected given the topic. In addition, spam is detected in two of the streams used in our tests, and not in the others, and we were able to verify, by examining the messages, that those streams indeed contained spam messages.

The rest of the paper is organized as follows. In Section \ref{sec:data}, the data acquisition system is described. Section \ref{sec:rd} reviews the renewal theory while the empirical approximation of the renewal density is detailed in Section \ref{sec:rdest}. In Section \ref{sec:detection}, the spam detection and the correlation characterization methods are addressed, showing the experimental results in Section \ref{sec:results}.

\section{DATA ACQUISITION SYSTEM}
\label{sec:data}

We implemented a data acquisition system based on the Twitter API, which allows access to the public timeline with certain limitations: a) only the first 10 pages of results can be accessed, b) maximum of 100 results per page, and c) minimum time between requests. The request of the first ten pages is automatic with a sleep time between requests, which has to be small enough so that the amount of new posted tweets during this time does not reach the maximum collectable ($1,000$), and high enough to not exceed the maximum number of daily requests or the maximum request rate. 

As will be discussed next, empirical estimation of the renewal density requires measuring the time in between tweets. Each message contains a date field, from which the time of the posting can be obtained. However, this timing information is relatively coarse (the resolution is one second), so that for high rate message streams (more than one message per second) multiple tweets can be recorded as having the same time. In our analysis, these tweets are treated as if they were submitted at the same time, i.e., their inter-arrival time is zero. Due to limited time resolution, our empirical pdfs have "spikes'' at the origin. While this is clearly an artifact of low time resolution, we decided to include these measurements in our pdf estimation. In any case, the phenomena of interest occur a larger time scales that should not be affected by this lack of resolution (e.g., spam messages may be posted with periods of the order of tens of minutes, while interactions between users in an active stream may be in a scale of minutes). 

Downsampling can be used as in \cite{Sean}. A random number of events would be grouped together and the inter-arrival value would become the difference between the first and the last tweet, reducing the amount of data and thus, the processing complexity. Lower complexity can facilitate analysis of multiple streams simultaneously and in real-time. However, with downsampling a longer acquisition time would be necessary in order to observe a sufficient number of samples for the estimated histograms to converge \cite{Markovich}.

\section{RENEWAL THEORY}
\label{sec:rd}

Let $M_1, M_2, \dotsc, M_n$ be positive, independent and identically distributed random variables (denote $M$ a generic random variable of the sequence), representing the inter-arrival time. The received time data format consists in the date and time (hours, minutes and seconds) of the event. Thus, a conversion to seconds followed by the calculation of the time difference between tweets is needed in order to get the relative time used for the renewal density. As a result, every tweet becomes a renewal process arrival. The partial sum $S_n= \sum_{i=0}^{n} M_i=M_1+M_2+\dotsb+M_n=S_{n-1}+M_n \quad \text{where} \quad S_0=0$, reflects the elapsed time since the beginning to the $n$th event. Finally, the renewal process $\{N(t); t>0\}$ represents the number of arrivals to the system in the interval $(0,t]$.

Let $f_{M_n}(t)$ be the pdf of the random variable $M_n$. As the partial sum $S_n$ is the sum of $n$ independent random variables, the pdf of the $n$th partial sum $f_{S_n}$ is defined as the convolution of the pdfs of first $n$ random variables which are also identically distributed. Thus, the pdf $f_{S_n}$ is given by the n-fold convolution of the pdf of $M_n$
\begin{equation}\label{fn}
	f_{S_n}(t)=f^{\ast n}(t)=(f_M\ast f_M\ast \dotsb f_M)(t).
\end{equation}
The pdf of the partial sum describes the probability that the $n$th event occurs at a certain time $t$. The renewal density is obtained by taking into account not only the $n$th event, but also all the arrivals. Hence, the probability of an arrival at $t$ independently of its order, would be the sum of the probabilities of all the possible arrivals, which leads to \cite{Sean}:
\begin{equation}
	r(t)=\sum_{n=0}^{\infty} f_{S_n}(t).
\label{r2}
\end{equation}

Furthermore, it is known that for a Poisson process with mean inter arrival time $\lambda$, the renewal density is given by $r(t)=1/\lambda$ \cite{Sean}. Thus, a non-constant rd means the process does not follow a Poisson distribution, i.e., there is some memory in the sequence of inter-arrival times. This difference will be used in Section~\ref{sec:detection} to characterize the background traffic.

\section{RENEWAL DENSITY ESTIMATION}
\label{sec:rdest}
A non-parametric estimation of the rd from empirical data is necessary as no information about the distribution of the arrivals is known. It is important to note that only one realization of the data can be obtained. In \cite{Markovich}, an estimate of the renewal function is proposed that uses no prior information about the form of the underlying distribution and is obtained from a single realization. The resulting nonparametric estimate is related to a histogram-type estimate where the unknown probabilities $Pr\{S_n \leq t\}$ are replaced by the corresponding empirical distribution functions and a limited number of terms $k$ is used in the summation. 

The method in \cite{Markovich} divides a realization with $m$ events into groups of $k$ terms with $k\leq m$. It starts with the first event $M_0$ and gets the partial sums of the first $k$ elements $\{S_1^0=M_1, S_2^0=S_1^0+M_2, \dotsc, S_k^0=S_{k-1}^0+M_k\}$. After that, the same process is followed, changing the starting point to the following event $M_1$ obtaining the sequence  $\{S_1^1, S_2^1, \dotsc, S_k^1\}$. This is repeated until the last element $M_k$ is reached. In other words, a sliding window filters a subset of events of the complete realization, obtaining the partial sums of the events in the window. As a result, a group of $m-k$ partial sums of the same order $j$ ($S_j^i$) is returned for $j = 1,\dotsc,k$. 

The estimation of the pdf $\widetilde{f}_{S_n}$ is computed non-parametrically using normalized histograms that tabulate the data from each partial sum realization. Each sequence of partial sums only contribute to the estimation of the respective order, i.e., $S_j^i$ contributes to $f_i$ only \cite{Sean}. Following (\ref{r2}), the resulting renewal density estimation $\widetilde{r}(t)$ is obtained by summing all the estimated normalized histograms. 

The estimation quality depends on the amount of data obtained, hence, it is possible to get a noisy rd unless we carefully select the bin size. In \cite{Shimazaki}, the optimal bin size is obtained by minimizing the cost function $C(\Delta)=(2k-v)/\Delta^2$ where $k=\frac{1}{N}\sum_{i=1}^Nk_i$ is the average number of events falling in the $i$th bin ($k_i$), $v=\frac{1}{N}\sum_{i=1}^N(k-k_i)^2$ is the variance and $\Delta$ corresponds to the bin size.

We also make use of a second estimation of the renewal density $\widetilde{r}'(t)$, obtained using (\ref{fn}) and (\ref{r2}). The different pdfs $\widetilde{f'}_{S_n}$ are computed through the $n$-fold convolution of the pdf $\widetilde{f}_{S_1}$ that is estimated empirically following the previous method. Thus, this renewal density only uses first order inter-arrival information. Hence, it is possible to consider  this rd as an approximation for orders greater than one, under the assumption that message arrivals are memoryless. This will help us determine how correlated is the acquired data. Note that its shape is similar to that of the rd obtained by a Poisson distribution: it is constant until it decreases abruptly due to the highest order pdf's tail shape.

In both estimations, the maximum inter-arrival time taken into account depends on the maximum pdf order used and the data rate. The lower the rate/higher the order, the higher maximum inter-arrival time. If any time value needs to be compared between rds estimated using different data, it is necessary to normalize it converting the units of seconds into tweets using the data rate factor.

\section{TWITTER TRAFFIC ANALYSIS}
\label{sec:detection}

\subsection{Periodic event detection}
Detection of periodic events is based on a Pearson's Chi-Square test \cite{Sean}. First, the empirical rd $\widetilde{r}(t)$ is obtained and normalized (this is denoted $\widetilde{r}_N(t)$), so that the maximum value is $10$, which allows a good performance of the test. Afterwards, $\widetilde{r}_N(t)$ is divided in sub-densities $\widetilde{r}_s(t)$, each of length $N_{bins}$. The smooth version $\widetilde{s}_s(t)$ is calculated with the trimmed mean method for each sub-density. The value of $\widetilde{s}_s(t)$ at a given $t$ is computed using only the $+/-T$ neighboring of $\widetilde{r}_N(t)$, i.e., \{$\widetilde{r}_N(t-T),\dotsc, \widetilde{r}_(t-1),\widetilde{r}_N(t+1), \dotsc, \widetilde{r}_N(t+T)$\} (note that $\widetilde{r}_N(t)$ is not used). Next, the values are ordered and the $T_{rim}\%$ top and bottom values are removed in order to determine $\widetilde{s}_s(t)$ at $t$ with the mean of the remaining values. Typically, $T_{rim}$ is selected to be $5\%-25\%$.

The collected traffic does not follow a Poisson distribution and consequently, fluctuations in the rd histogram may appear, which could be detected as false positives if $N_{bins}$ is too long. Thus, as the length of each $\widetilde{r}(t)$ varies significantly, a number of sub-densities is fixed in order to adapt $N_{bins}$ to each histogram.

The value $\chi^2_i$ for the $i$th sub-density is computed using
\begin{equation}
 \chi^2_i=\sum_{m=1}^{N_{bins}}\frac{(\widetilde{r}_{s(i)}(m)-\widetilde{s}_{s(i)}(m))^2}{\widetilde{s}_{s(i)}(m)},
\label{chi}
\end{equation}
 which is used to compute the Pearson statistic, $p$, by evaluating the cumulative density function of the Chi-Square Distribution with $N_{bins}$ degrees of freedom at $\chi^2_i$  \cite{Sean}. Finally, if $p>1-P_{FA}$, where $P_{FA}$ is the probability of false alarm, an anomaly is detected. Typical values of $P_{FA}$ are $0.01$ or $0.05$.
 
\subsection{Background traffic characterization}
\begin{figure}[htb]
\begin{minipage}[b]{.48\linewidth}
  \centering
  \centerline{\includegraphics[width=4.3cm]{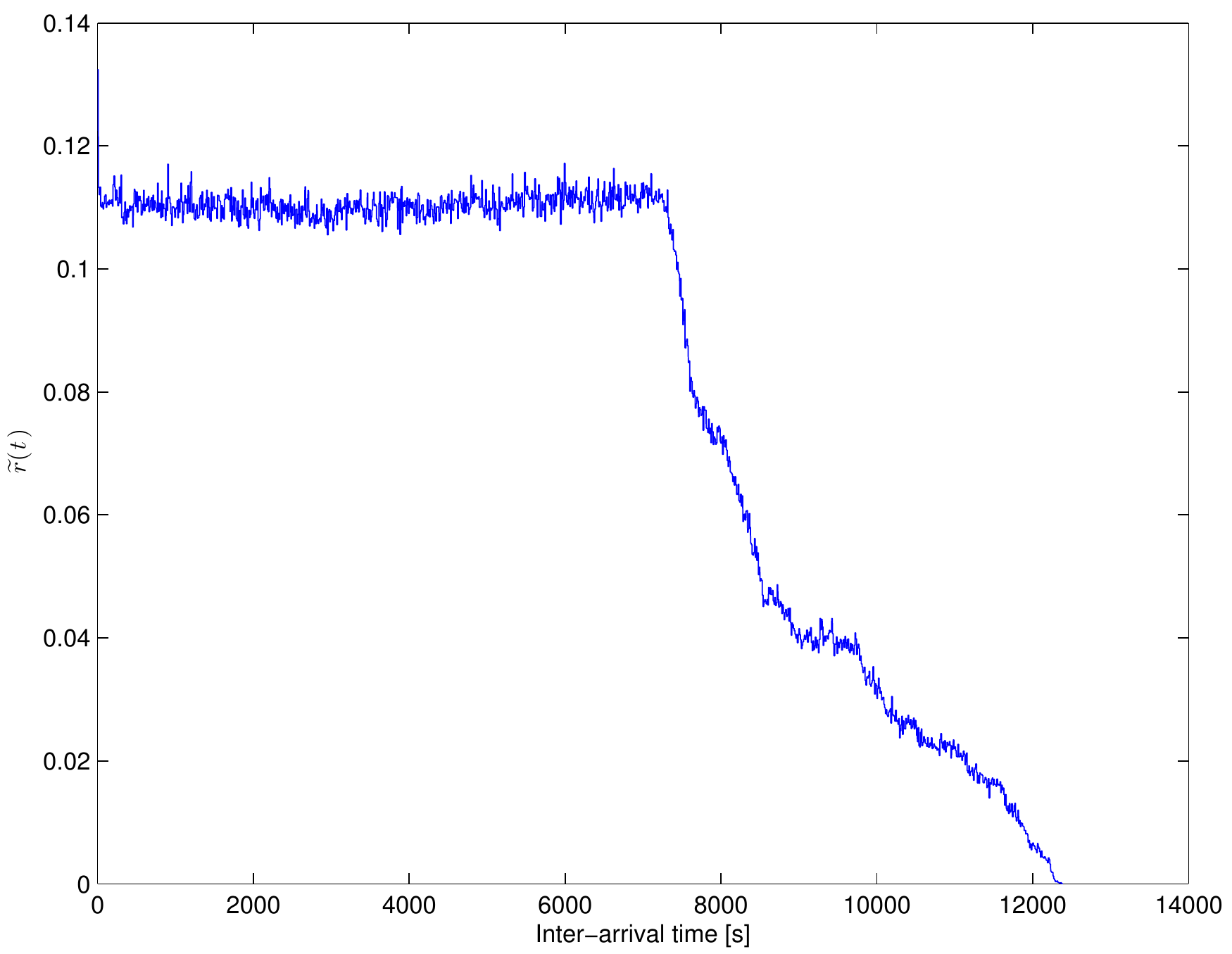}}
  \centerline{(a) Empirical rd $\widetilde{r}(t)$}\medskip
\end{minipage}
\hfill
\begin{minipage}[b]{0.48\linewidth}
  \centering
  \centerline{\includegraphics[width=4.3cm]{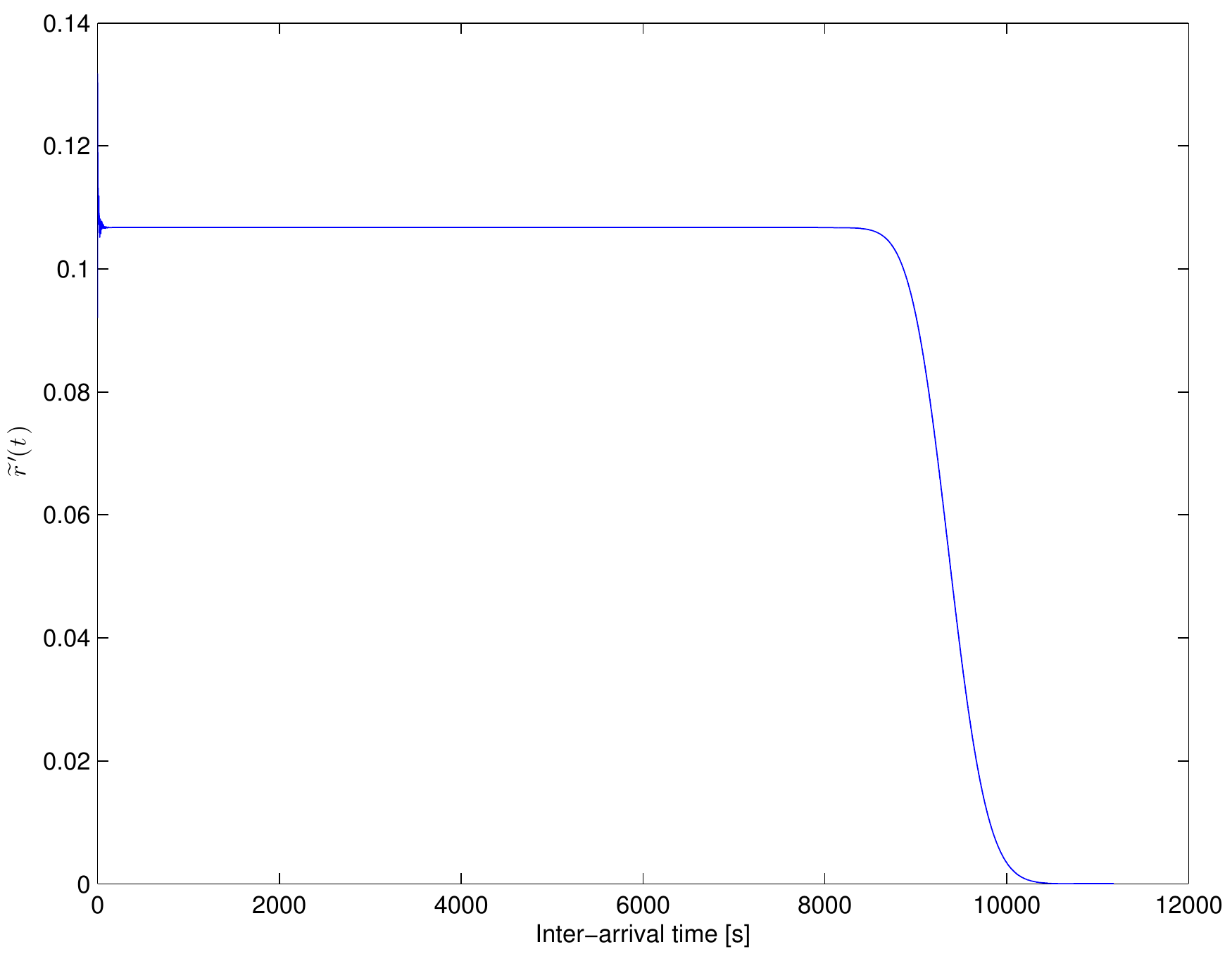}}
  \centerline{(b) Convolution rd $\widetilde{r}'(t)$ }\medskip
\end{minipage}
\begin{minipage}[b]{.48\linewidth}
  \centering
  \centerline{\includegraphics[width=4.3cm]{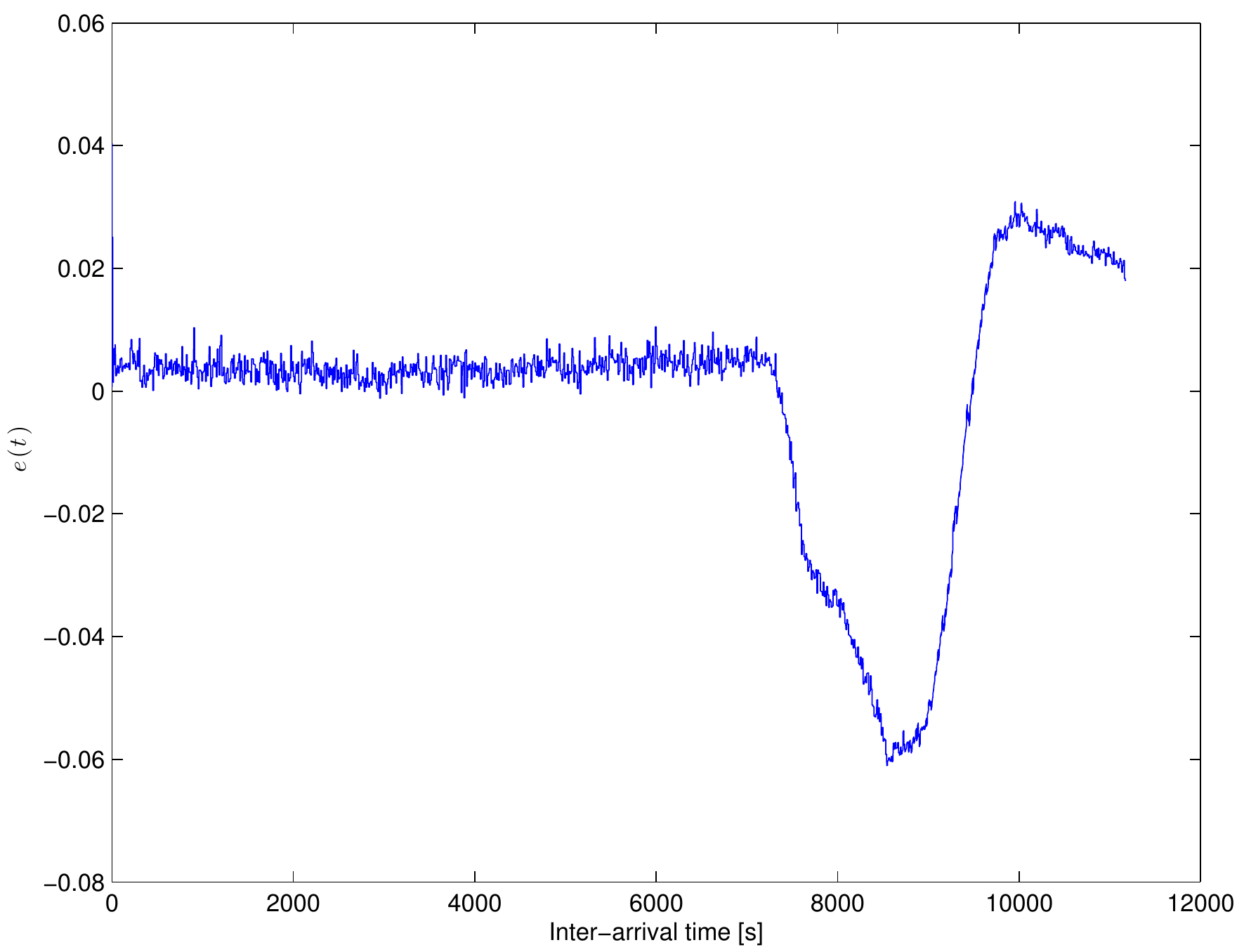}}
  \centerline{(c) Difference $e(t)$}\medskip
\end{minipage}
\hfill
\begin{minipage}[b]{0.48\linewidth}
  \centering
  \centerline{\includegraphics[width=4.3cm]{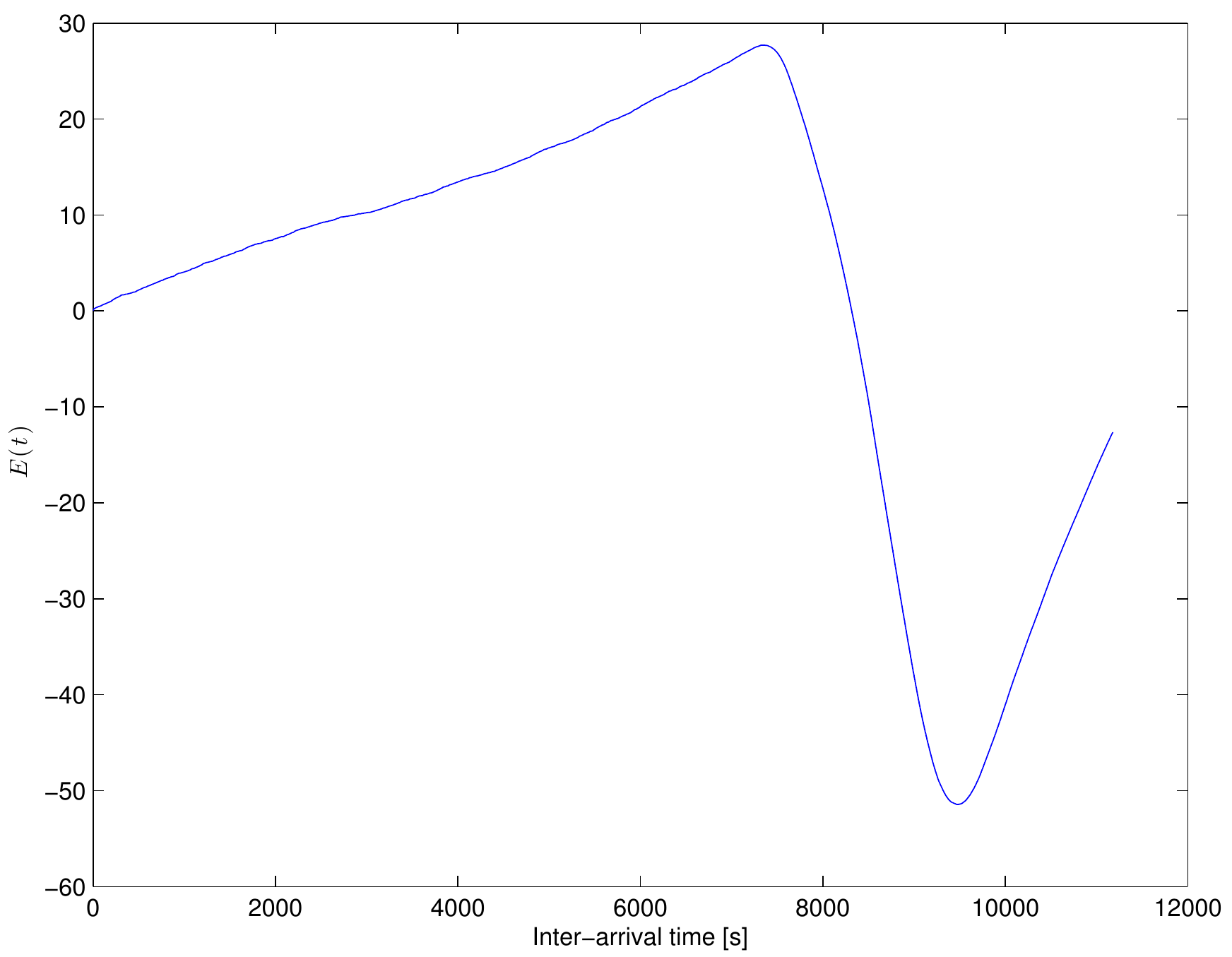}}
  \centerline{(d) Cumulative difference $E(t)$ }\medskip
\end{minipage}
\vspace{-.45cm}
\caption{$\widetilde{r}(t)$, $\widetilde{r}'(t)$, $e(t)$ and $E(t)$ of traffic with the keyword \emph{surf} and maximum pdf order estimated $1,000$.}
\label{fig:surf}
\end{figure}

In order to characterize the background traffic, which corresponds to the stream without periodic events, both estimations of the rd are used. In general, the values for low inter-arrival times in $\widetilde{r}(t)$ are higher than in $\widetilde{r}'(t)$, as can be seen in the example of Figure~\ref{fig:surf}(a) and \ref{fig:surf}(b). This means that using only first order inter-arrivals underestimates the number of tweets for the lower inter-arrival times. As the time increases, the empirical estimation starts decreasing before the estimation based on convolution. 
Letting $e(t)=\widetilde{r}(t)$-$\widetilde{r}'(t)$ be the difference between the empirical estimation and the convolution estimation of the rd, 
we can see  (Figure~\ref{fig:surf}(c)) that areas where $e(t)$ is flat and close to zero correspond to time scales at which the data is less correlated and thus has behavior close to that of a memoryless process (as estimated using the convolution). 

The main difference between $e(t)$ of various data sets lies on the lower part of the time axis. The rds with more random traffic, have a more constant beginning, e.g., Figure~\ref{fig:surf}(c). In order to differentiate the different starting cases, the cumulative difference $E(t)=\sum_{k=0}^t\widetilde{r}(k)-\widetilde{r}'(k)$ is obtained (Figure~\ref{fig:surf}(d)). The maximum of $E(t)$ is used to differentiate the different types of message streams, as this maximum measures the area under $e(t)$ from the origin until the point where the error becomes negative. Thus, the higher the maximum value is, the more correlated the data is (more error between a memoryless estimate and the empirically measured rd). This is justified by the fact that the convolution estimation of the rd does not have information of the higher orders and both estimations have the same first order pdf. It is important to note that the maximum value of $E(t)$ needs to be normalized by the number of normalized pdfs used, which may vary along the different data sets.

\section{EXPERIMENTAL RESULTS}
\label{sec:results}

\begin{figure}[htb]
\begin{minipage}[b]{1.0\linewidth}
  \centering
 \centerline{\includegraphics[width=8.5cm]{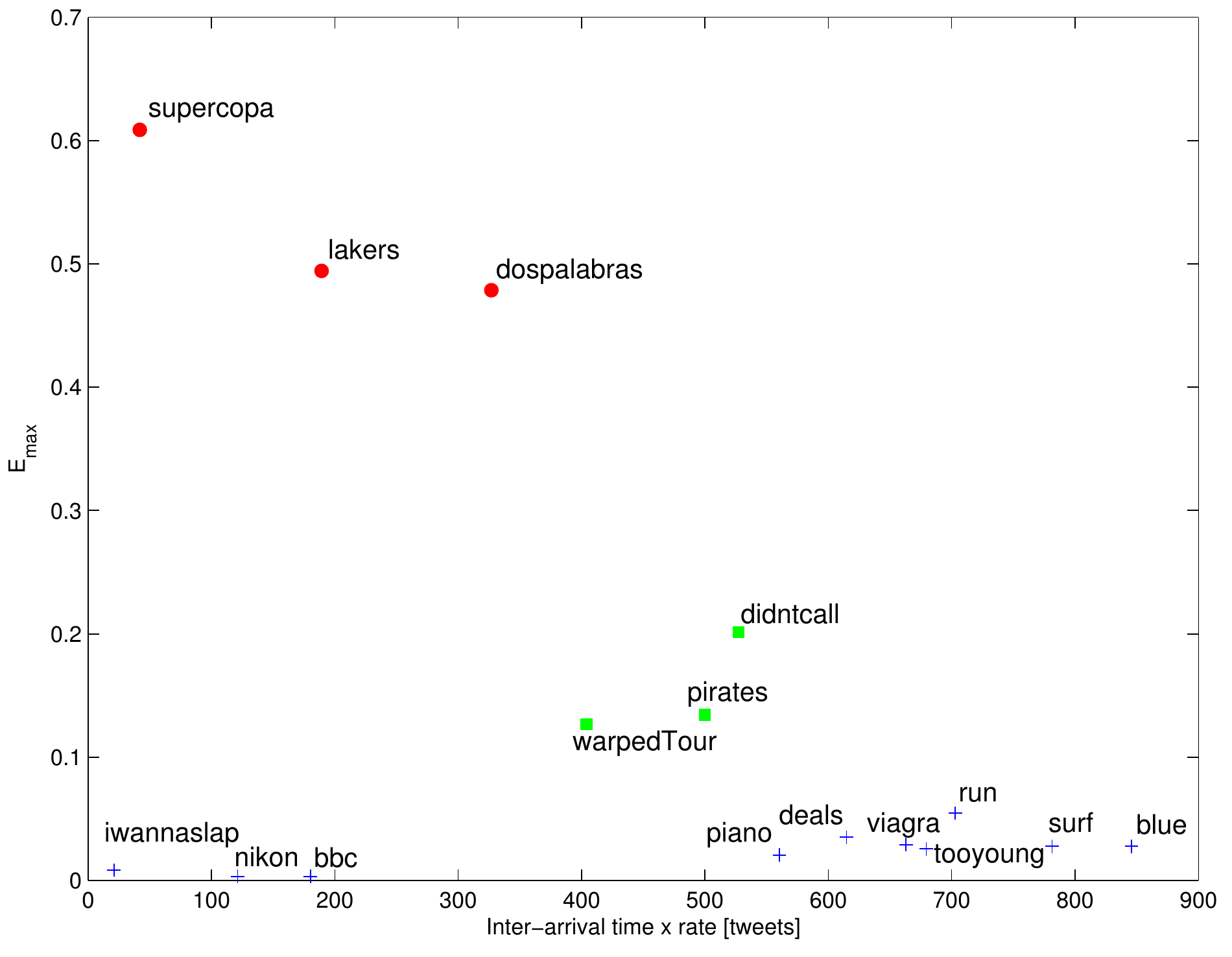}}
\end{minipage}
\vspace{-.7cm}
\caption{Representation of the maximum value of $E(t)$ versus the normalized position for each keyword}
\label{fig:max}
\vspace{-0.2cm}
\end{figure}

We collect data from Twitter with three different types of keywords. The first one are general words that can be used in multiple topics. The second group consists of keywords related to social events and the third one are trend topics. In Table~\ref{tab:keywords}, the classification of the used keywords according to their type is shown. Messages in the 'Events' category were collected during the event, as well as immediately following it.  For each of those keywords, $\widetilde{r}(t)$ and $\widetilde{r}'(t)$ are calculated. The bin size of the empirical version is optimized as explained in Section~\ref{sec:rdest}.  Next, the difference between both estimations, $e(t)$, and the cumulative error, $E(t)$, are calculated and used to determine the memory in the sequence of inter-arrival times.  Figure~\ref{fig:max}, shows for each of the keywords the peak value of $E(t)$ and its position, which is normalized by the average data rate. Thus a position of, for example, 100 in the horizontal axis, correspond to the time it takes on average for 100 tweets to be sent in that stream.

\begin{table}[htdp]
\begin{center}
\begin{tabular}{|c|c|c|}
\hline
General & Events & Trend topics \\ 
\hline
bbc & lakers & didntcall \\
blue & supercopa & dospalabras \\
deals & warpedTour & iwannaslap \\
nikon & pirates & toogoung\\
piano &  & \\
run &  & \\
surf &  & \\
viagra &  & \\
\hline
\end{tabular}
\caption{Keywords classification.}
\end{center}
\label{tab:keywords}
\end{table}%

With those results, three zones can be differentiated with ease. The lower value ones ('+' symbol) correspond to the more random traffic; the middle zone (squares) have some more correlation while the upper zone (circles) is for the more correlated ones. The 'General' type have a memoryless behavior, which was expected as each user posts a tweet containing the keyword in a wide range of different topics. In contrast, keywords associated with events have a higher correlation as they are influenced by other messages in the stream as well as specific occurrences within the event (for example, multiple messages may be sent if a goal is scored in a soccer game). However, \emph{warpedTour} and \emph{pirates} are in the middle zone. This is due to the fact that their rate is very low, which indicates that the posted tweets are not very dependent on each other. Finally, the 'Trend topics' type are distributed in all the zones. Analyzing the text and users of the tweets, it is possible to find thematic bursts in the correlated ones, while tweets seem independent in the rest.

\begin{figure}[htb]
\begin{minipage}[b]{1.0\linewidth}
  \centering
 \centerline{\includegraphics[width=8.5cm]{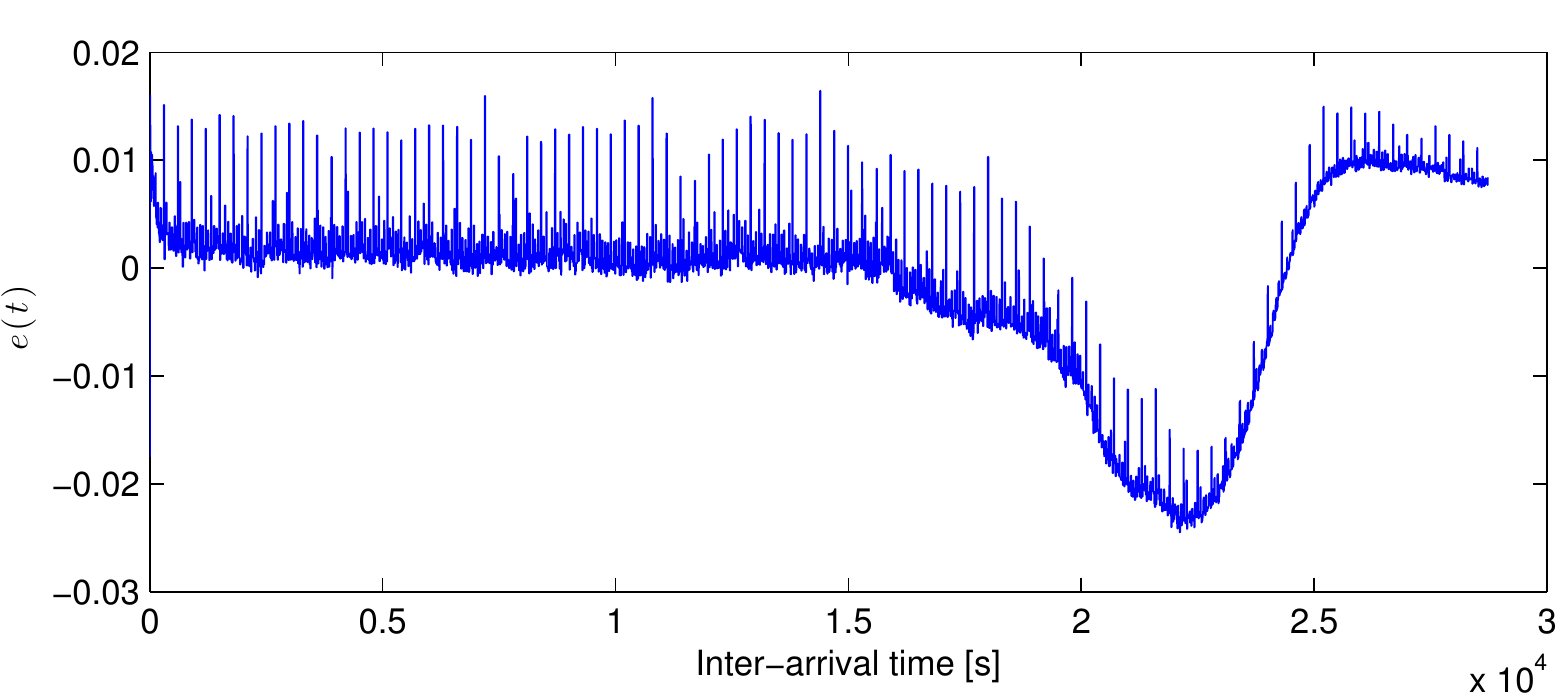}}
\end{minipage}
\vspace{-0.65cm}
\caption{Difference $e(t)$ of keyword \emph{viagra} with overlaying spam.}
\label{fig:spam}
\vspace{-0.15cm}
\end{figure}

For the periodic event detection, only keywords with over $5000$ tweets have been tested to ensure the convergence of the histogram. We select $T=N_{bins}/2$ in order to compute the trimmed mean in an interval equal to the sub-density and $T_{rim}=35\%$ as it gives a more robust behavior \cite{Sean}. Some keywords are widely used by spammers, e.g., \emph{viagra}, i.e., spam appears overlying the background traffic. Figure~\ref{fig:spam} shows the empirical rd estimation with this periodic traffic, obtained with the mentioned keyword.
Note that in some cases (e.g., as in Fig.~\ref{fig:spam}) the volume of spam is so significant that it is easy to detect. However, our method can detect the presence of spam even in cases where the volume of spam messages is relatively low. For example our algorithm detected the presence of spam traffic under the keyword \emph{nikon}, which could clearly be seen from observing the rd plot, but was confirmed by observing the messages' text. No periodic events were detected for the other keywords.

\section{CONCLUSION}
\label{sec:conclusion}
In this paper we approach the information extraction from Twitter's traffic about the memory in the sequence of inter-arrival times using renewal theory, and use this also to detect periodic events in the stream. Three types of keywords have been used in order to obtain different testing scenarios and try the method presented. 
With all, we conclude that it is possible to classify a stream in one of three different zones with different correlation grades and detect periodic events using Pearson's Chi-Square test, which in most of the cases are related to spam.

\bibliographystyle{IEEEbib}
\bibliography{bib}

\end{document}